\documentclass[12pt]{article}
\usepackage{jhep-mod}
\usepackage{bm}
\usepackage{amssymb,amsmath,amsthm}
\usepackage{mathrsfs}
\usepackage{xcolor}
\usepackage{amssymb}% http://ctan.org/pkg/amssymb
\usepackage{tikz}
\definecolor{purple}{rgb}{1,0,1}
\definecolor{lime}{HTML}{A6CE39} % needs xcolor
\definecolor{lime}{HTML}{A6CE39}
\newcommand{\orcidicon}{%
	\begin{tikzpicture}
	\draw[lime, fill=lime] (0,0) 
		circle [radius=0.16] 
		node[white] {{\fontfamily{qag}\selectfont \tiny ID}};
	\draw[white, fill=white] (-0.0625,0.095) 
		circle [radius=0.007];
	\end{tikzpicture}
	\hspace{-5mm}
}
\newcommand\orcidMatt{{\href{https://orcid.org/0000-0003-1088-6485}{\orcidicon}}}
\definecolor{purple}{rgb}{1,0,1}

\DeclareMathOperator{\tr}{tr}
\begin{document}
%=================================================================
%=================================================================
\title{Regularization versus renormalization: \\
\leftline{Why are Casimir energy differences so often finite?}}

\author{Matt Visser\orcidMatt{}}

\affiliation{%
School of Mathematics and Statistics,
Victoria University of Wellington, \\PO Box 600, Wellington~6140,~New~Zealand.
}

\abstract{
One of the very first applications of the quantum field theoretic vacuum state was in the development of the notion of Casimir energy. Now field theoretic Casimir energies, considered individually, are always infinite. 
But \emph{differences} in Casimir energies (at worst regularized, not renormalized) are quite often finite --- a fortunate circumstance which luckily made some of the early calculations, (for instance, for parallel plates and hollow spheres), tolerably tractable. 
We shall explore the extent to which this observation can be made systematic.
For instance: What are necessary and sufficient conditions for Casimir energy \emph{differences} to be finite (with regularization but without renormalization)?  
And, when the Casimir energy \emph{differences} are not formally finite, can anything useful nevertheless be said by invoking renormalization? We shall see that it is the \emph{difference} in the first few Seeley--DeWitt coefficients that is central to answering these questions. In particular, for any collection of conductors (be they perfect or imperfect) and/or dielectrics, as long as one merely moves them around without changing their shape or volume, then physically the  Casimir energy difference (and so also the physically interesting Casimir forces) are guaranteed to be finite without invoking any renormalization. 

\bigskip
\noindent
{\sc Keywords:} {Casimir energy; quantum field theory; renormalization; regularization; \\
finiteness; heat kernel expansion; Seeley--DeWitt coefficients.}

\bigskip
\noindent
{\sc Pacs:} {03.70.+k; 11.10.-z; 12.20.-m}

\bigskip
\noindent
{\sc Date:}  7 January 2016; 7 January 2019; \LaTeX-ed \today

\bigskip
\noindent
{\sc arXiv:}  1601.01374 [quant-ph]

\bigskip
\noindent
{\sc Published:} Particles  {\bf 2 \# 1} (2019)  14-31. 
}
%=================================================================
\maketitle
%=================================================================
\def\d{{\mathrm{d}}}
\def\I{{\mathcal{I}}}
\def\O{{\mathcal{O}}}
\def\Im{{\mathrm{Im}}}
\def\implies{\Rightarrow}
\def\half{{1\over2}}
%----------------------------------------------------------------
\def\erf{{\mathrm{erf}}}
\def\erfc{{\mathrm{erfc}}}
%----------------------------------------------------------------
\def\HRULE{{\bigskip\hrule\bigskip}}
%----------------------------------------------------------------
%----------------------------------------------------------------
\def\SIZE{1.00}
%----------------------------------------------------------------
%-------------------------------------------------------------------------------------------------------------------------------------------
\def\lint{\hbox{\Large $\displaystyle\int$}}   %needs \usepackage{amssymb} [large integral]
\def\hint{\hbox{\huge $\displaystyle\int$}}  %needs \usepackage{amssymb} [huge integral]
%------------------------------------------------------------------------------------------------------------------------------------------
\newcommand{\V}{\mathbf{V}}
\newtheorem{lemma}{Lemma}
\newtheorem{theorem}{Theorem}
%---------------------------------------------------------------------------------------------------------------------------------------------
%---------------------------------------------------------------------
\setlength\fboxsep{-3.0pt}
\setlength\fboxrule{6.0pt}
%--------------------------------------------------------------------
%%========================================================
%%========================================================

%---------------------------------------------------------------------------------------------------------------------------------------------
\section{Introduction}
%---------------------------------------------------------------------------------------------------------------------------------------------
\label{S:intro}
%---------------------------------------------------------------------------------------------------------------------------------------------
Quantum field theoretic Casimir energies (considered in isolation) are typically infinite, requiring both regularization and renormalization to extract mathematically sensible answers, this at the cost of sometimes obscuring the underlying physics~\cite{Casimir:1948, Boyer:1968, DeWitt:1975, Plunien:1986, Blau:1988, Blau:1988b, Visser:2002, Bordag:2001}. 
On the other hand Casimir energy \emph{differences} are quite often finite, and have a much more direct physical interpretation~\cite{Casimir:1948, Boyer:1968}.  That is, Casimir energy differences often merely require regularization, not renormalization. Additional background and general developments may be found in references~\cite{Balian:1977,  Kay:1978, Bordag:1983, Graham:2002x, Kenneth:2002, Fulling:2003, Gies:2003, Jaffe:2003, Kenneth:2006, Emig:2007, Kenneth:2007, Klich:2009, Rahi:2009, Klimchitskaya:2009, Abalo:2012, Fulling:2018}. 
In this article, I shall first argue (mathematically) that there are a large number of interesting physical situations where the Casimir energy differences,
(and so the Casimir energy forces), are automatically known to be finite, even before starting specific computations. 
Secondly, I shall argue (mathematically) that one can often develop physically interesting ``reference models'' such that the Casimir energy difference between the physical system and the ``reference model'' is known to be finite, 
even before starting specific computations. 
(I~will not actually calculate any Casimir energies --- knowing that the result you are after is finite, and so mathematically and physically well-defined, is often more than half the battle.) Finally I shall consider a general class of physically reasonable cutoff functions that give useful physical information even when energy differences are formally infinite; the cutoff is carefully defined so that one can invoke the ordinary universal heat-kernel expansion, instead of having to resort to the less well understood and non-universal cylinder-kernel (Poisson kernel) expansion for pseudo-differential operators~\cite{Fulling:2003}.

I shall first start with a simple formal mathematical argument to get the discussion oriented, and then provide a more careful mathematical argument in terms of regularized (but not renormalized) Casimir energies.  Since the entire goal of the article is to make physical sense of differences in mathematically divergent series of the form $E = {1\over2} \hbar \sum_n \omega_n$ there will unavoidably be some significant mathematical preliminaries ---
but ultimately we shall be very much focussed on the underlying physics. 

%=================================================================
\section{Formal mathematical argument}
%=================================================================
The formal argument starts with the elementary fact that 
 for any two numbers $\omega$ and $\omega_*$ their difference can be represented by the convergent integral:
\begin{equation}
\omega-\omega_* = {1\over\sqrt{4\pi}} \int_0^\infty {\d s\over s^{3/2}} \left\{ e^{-s\omega_*^2 } - e^{-s\omega^2} \right\}.
\end{equation}
(Note that it is essential to perform the subtraction before the integration. Otherwise the two integrals are individually infinite.)
To establish this identity first integrate by parts
\begin{equation}
{1\over\sqrt{4\pi}} \int_0^\infty {\d s\over s^{3/2}} \left\{ e^{-s\omega_*^2 } - e^{-s\omega^2} \right\}
= - {1\over\sqrt{\pi}} \int_0^\infty {\d s\over s^{1/2}} \left\{ \omega_*^2 e^{-s\omega_*^2 } - \omega^2 e^{-s\omega^2} \right\},
\end{equation}
and note that the two integrals on the right now individually converge. Finally use the identity
\begin{equation}
\int_0^\infty {\d s\over \sqrt{s}} \; e^{-s \omega^2} = {\sqrt{\pi}\over \omega},
\end{equation}
and the result is established.

Now let $\omega_n$ and $(\omega_*)_n$ be two infinite sequences of numbers, then we can certainly compute the sum of differences for the first $M$ terms:
\begin{equation}
\sum_{n=1}^M \{\omega_n-(\omega_*)_n\} = 
{1\over\sqrt{4\pi}} \sum_{n=1}^M \int_0^\infty {\d s\over s^{3/2}} \left\{ e^{-s(\omega_*^2)_n} - e^{-s\omega_n^2} \right\}.
\end{equation}
Now taking $M\to\infty$ and formally interchanging integral and summation, 
(\emph{and I will justify this rather cavalier step much more carefully later on}):
\begin{equation}
\sum_n \{\omega_n-(\omega_*)_n\} = 
{1\over\sqrt{4\pi}} \int_0^\infty {\d s\over s^{3/2}}  \sum_n \left\{ e^{-s(\omega_*)^2_n } - e^{-s\omega_n^2} \right\}.
\end{equation}
Then in terms of the heat kernels  $K(s)$ and $K_*(s)$ defined by
\begin{equation}
K(s) = \sum_n e^{-s\omega_n^2 }; \qquad K_*(t) = \sum_n e^{-s(\omega_*)_n^2},
\end{equation}
we formally have:
\begin{equation}
\sum_n \{\omega_n-(\omega_*)_n\} = 
 \int_0^\infty {\d s\over s} {1\over\sqrt{4\pi s}} \left\{ K_*(s) - K(s)\right\}.
\end{equation}
It is well-known that possible high-energy (ultra-violet, UV) divergences in this integral are related to the small-$s$ behaviour of the integrand.
Possible low-energy (infra-red, IR) divergences in this integral are related to the large-$s$ behaviour of the integrand, and are for our current purposes uninteresting;  possible IR divergences are easily fixed by putting the system in a large but finite box.
Indeed, (\emph{now assuming that the $\omega_n^2$ and $(\omega_*)_n^2$ are in fact 
the eigenvalues of some second-order linear differential operators}),  then
by invoking the standard Seeley--DeWitt asymptotic expansion (the heat kernel expansion) we have both
\begin{equation}
K(s) = (4\pi s)^{-d/2}  \left\{ \sum_{i=0}^N  a_{i/2} \; s^{i/2} + \O\left(s^{(N+1)/2}\right) \right\},
\end{equation}
and
\begin{equation}
K_*(s) = (4\pi s)^{-d/2}  \left\{ \sum_{i=0}^N  (a_*)_{i/2} \; s^{i/2} + \O\left(s^{(N+1)/2}\right) \right\}.
\end{equation}
Note that $d$ is the number of space dimensions. As will be discussed more fully below, the integer indexed coefficients $a_n$ have both bulk and boundary contributions, 
while the half-integer indexed coefficients $a_{n+{1\over2}}$ have only boundary contributions.
(Some references re-label the Seeley--DeWitt coefficients $a_n\to a_{2n}$ to force all the subscripts to be integer. 
I prefer to keep the integer/half-integer distinction manifest because there are significant qualitative differences between these coefficients.)
Then for the difference in heat kernels we have:
\begin{equation}
K_*(s) -K(s)= (4\pi s)^{-d/2}  \left\{ \sum_{i=0}^N  \left\{ (a_*)_{i/2}  - a_{i/2}   \right\} \; s^{i/2} + \O\left(s^{(N+1)/2}\right) \right\}.
\end{equation}

Now choose $N=d+1$, and cut off the $s$-integration at some convenient but arbitrary upper limit $S_*$, then formally
\begin{equation}
\sum_n \{\omega_n-(\omega_*)_n\} = 
 \int_0^{S_*} {\d s\over s}  (4\pi s)^{-(d+1)/2}  \left\{ \sum_{i=0}^{d+1}  \left\{ (a_*)_{i/2}  - a_{i/2}   \right\} \; s^{i/2} \right\}
+(\hbox{UV finite}).
\end{equation}
Here the designation ``UV finite'' means that any remaining terms contributing to the ``UV finite'' piece are now guaranteed to not have any infinities coming from the $s\to0$ region of integration. 
That is, taking $E_\mathrm{Casimir} = {1\over2} \hbar \sum_n \omega_n$, and defining $\Delta a_i = a_i - (a_*)_i$, we have the formal result:
\begin{equation}
\Delta (\hbox{Casimir Energy)}  = - 
{\hbar\over2} \int_0^{S_*} {\d s\over s}  (4\pi s)^{-(d+1)/2}  \left\{ \sum_{i=0}^{d+1}   \Delta a_{i/2} \; s^{i/2} \right\}
+(\hbox{UV finite}).
\end{equation}
All of the potentially UV-divergent terms are now concentrated in the $d+2$ leading terms proportional to the $\Delta a_i$.
The rest of this article will involve several powerful refinements on this simple theme.
Note that for finiteness of the Casimir energy difference:
\begin{itemize}
\itemsep-1pt
\item In 3+1 dimensions we would want $\Delta a_0=\Delta a_{1/2} = \Delta a_1 = \Delta a_{3/2} = \Delta a_2 = 0$.
\item In 2+1 dimensions we would want $\Delta a_0=\Delta a_{1/2} = \Delta a_1 = \Delta a_{3/2}  = 0$.
\item In 1+1 dimensions we would want $\Delta a_0=\Delta a_{1/2} = \Delta a_1 = 0$.
\item In 0+1 dimensions we would want $\Delta a_0=\Delta a_{1/2} = 0$. 
\end{itemize}
Generally, in $d$ space dimensions, if we are comparing any two physical systems for which the first $d+2$ Seeley--DeWitt coefficients are equal, then the difference in Casimir energies will be finite. (That is, finite after regularization, subtraction and removal of the regulator --- no renormalization is required.)

%=================================================================
\section{Exact mathematical argument}
%=================================================================
Let us now regularize everything a little more carefully, to develop an exact and mathematically rigorous rather than formal argument. Initially we shall use the complementary error function [$\erfc(x) = 1-\erf(x)$] as a particularly simple and mathematically transparent regulator, monotonically interpolating between $\erfc(0)=1$ and $\erfc(\infty)=0$, but will subsequently show that physically almost any smooth cutoff function will do. 
At a technical level the complementary error function regulator is particularly easy to relate to the standard Seeley-DeWitt expansion; in contrast exponential regulators such as $\exp(-x)$ lead to pseudo-differential kernels such as $\exp(-s\sqrt{\Delta})$ and cylinder expansions (Poisson expansions) which are technically much messier to deal with --- extra non-universal terms show up in the cylinder (Poisson) expansion, see reference~\cite{Fulling:2003}.
We start from the exact result that:
\begin{equation}
\Omega \; e^{-\omega^2/\Omega^2} = - \int_{\Omega^{-2}}^\infty {\d\over\d s} \left( s^{-1/2} e^{-s\omega^2} \right) \, \d s,
\end{equation}
and note that this implies
\begin{equation}
\omega \; \erfc(\omega/\Omega) = {\Omega\over\sqrt{\pi}} \; e^{-\omega^2/\Omega^2} 
- {1\over\sqrt{4\pi}} \int_{\Omega^{-2}}^\infty {\d s\over s^{3/2} } \; e^{-s\omega^2}\;.
\end{equation}
Summing over the eigen-frequencies, this leads to the further exact result that:
\begin{equation}
\sum_n \omega_n \; \erfc(\omega_n/\Omega) = {\Omega\over\sqrt{\pi}} \sum_n e^{-\omega_n^2/\Omega^2} 
- {1\over\sqrt{4\pi}}\sum_n  \int_{\Omega^{-2}}^\infty {\d s\over s^{3/2} } \;e^{-s\omega_n^2}\;.
\end{equation}
But, because all the relevant quantities are now guaranteed finite, we can now safely exchange sum and integral to obtain the exact (no longer just formal) result:
\begin{equation}
\sum_n \omega_n \; \erfc(\omega_n/\Omega) = {\Omega\over\sqrt{\pi}} \sum_n e^{-\omega_n^2/\Omega^2} 
- {1\over\sqrt{4\pi}} \int_{\Omega^{-2}}^\infty {\d s\over s^{3/2} }\sum_n e^{-s \omega_n^2}\;.
\end{equation}
Then in terms of the heat kernel:
\begin{equation}
\sum_n \omega_n \; \erfc(\omega_n/\Omega) = {\Omega\over\sqrt{\pi}} \;K(\Omega^{-2})
- {1\over\sqrt{4\pi}} \int_{\Omega^{-2}}^\infty {\d s\over s^{3/2} }\; K(s).
\end{equation}
Now apply the Seeley--DeWitt asymptotic expansion:
\begin{equation}
K(s) = (4\pi s)^{-d/2}  \left\{ \sum_{i=0}^N  a_{i/2} \; s^{i/2} + \O\left(s^{(N+1)/2}\right) \right\}.
\end{equation}
But then for the heat kernel term, (choosing $N=d$\,),  we have:
\begin{equation}
{\Omega\over\sqrt{\pi}} K(\Omega^{-2}) = 
2\left(\Omega\over\sqrt{4\pi}\right)^{d+1}  \left\{ \sum_{i=0}^{d}   a_{i/2}   \; \Omega^{-i} \right\} +(\hbox{finite as } \Omega\to\infty).
\end{equation}
Working with the integral term is a little trickier.
In the integral we instead find it useful to {choose} $N=d+1$. Then, treating the logarithmic term separately, and cutting off the $s$-integration at some convenient finite but arbitrary value $S_*$, we have
\begin{eqnarray}
\int_{\Omega^{-2}}^\infty {\d s\over s}  {1\over\sqrt{4\pi s}}\, K(s) &=&
{1\over\sqrt{4\pi}} \int_{\Omega^{-2}}^{S_*} {\d s\over s^{3/2}}  (4\pi s)^{-d/2}  \left\{ \sum_{i=0}^{d}  \left\{ a_{i/2}   \right\} \; s^{i/2} \right\}
\nonumber\\
&&\vphantom{\Big{|}}
+{a_{(d+1)/2}\over(4\pi)^{(d+1)/2}} \ln(S_* \Omega^2) 
+(\hbox{finite as } \Omega\to\infty).
\end{eqnarray}
That the $a_{(d+1)/2}$ term leads to logarithmic term in the Casimir energy (and effective action) is well-known.  See for instance references~\cite{Blau:1988, Blau:1988b, Hochberg:1998, Visser:1994}.
Performing the remaining integrals, lumping the $S_*$ dependence into the finite piece, and introducing a convenient scale $\mu$, we have:
\begin{eqnarray}
\int_{\Omega^{-2}}^\infty {\d s\over s}  {1\over\sqrt{4\pi s}} \, K(s) &=&
- {1\over(4\pi)^{(d+1)/2}} \left\{ \sum_{i=0}^{d}  {a_{i/2}  \; \Omega^{d+1-i}\over (d+1-i)/2}\right\} 
+{a_{(d+1)/2}\over(4\pi)^{(d+1)/2}} \ln(\Omega^2/\mu^2) 
\nonumber\\
&&\qquad\qquad\vphantom{\Big{|}}
+(\hbox{finite as } \Omega\to\infty).
\end{eqnarray}
Now assembling all the pieces:
\begin{eqnarray}
\sum_n \omega_n \; \erfc(\omega_n/\Omega) &=&  2\left(\Omega\over\sqrt{4\pi}\right)^{d+1}  \left\{ \sum_{i=0}^{d}  \left\{ a_{i/2}   \right\} \; \Omega^{-i} \right\}
+ {2\over(4\pi)^{(d+1)/2}} \left\{ \sum_{i=0}^{d}  {a_{i/2}  \; \Omega^{d+1-i}\over d+1-i}\right\} 
\nonumber\\
&&\qquad\qquad\vphantom{\Bigg{|}}
+{a_{(d+1)/2}\over(4\pi)^{(d+1)/2}} \ln(\Omega^2/\mu^2) 
+(\hbox{finite as } \Omega\to\infty).
\end{eqnarray}
We now have the exact mathematical result: 
\begin{lemma}
For the eigen-frequencies $\omega_n^2$ arising from a second-order differential operator with Seeley-DeWitt coefficients $a_i$ one has:
\emph{
\begin{equation}
\sum_n \omega_n \; \erfc(\omega_n/\Omega) 
=  \left\{ \sum_{i=0}^{d}  k_i \; a_{i/2}  \; \Omega^{d+1-i} \right\}
+a_{(d+1)/2} \; \ln(\Omega^2/\mu^2) 
+(\hbox{finite as } \Omega\to\infty).\qquad
\end{equation}
}
\!Here $k_i = 2(1+{1\over d+1-i})$, but for our current purposes the specific values of the dimensionless coefficients $k_i$ are not at all important.
\hfill{$\Box$}
\end{lemma}
\noindent
We now wish to take this mathematical result and turn it into a physics statement about the Casimir energy.

%=================================================================
\section{Defining and using the erfc-regularized Casimir energy}
%=================================================================
Define the erfc-regularized Casimir energy as:
\begin{equation}
(\hbox{Casimir energy})_\mathrm{erfc} = {1\over2}\hbar \sum_n \omega_n \; \erfc(\omega_n/\Omega).
\end{equation}
This quantity is guaranteed to be finite as long as $\Omega$ is finite. 
Then:
\begin{eqnarray}
(\hbox{Casimir energy})_\mathrm{erfc} &=&
 {1\over2} \hbar\left\{ \sum_{i=0}^{d}  k_i \; a_{i/2}  \; \Omega^{d+1-i} \right\}
+  {1\over2} \hbar  \; a_{(d+1)/2} \; \ln(\Omega^2/\mu^2) 
\nonumber\\&&\vphantom{\Bigg|}
+(\hbox{finite as } \Omega\to\infty).
\end{eqnarray}
Now take differences:
\begin{eqnarray}
\Delta (\hbox{Casimir energy})_\mathrm{erfc}  &=&
 {1\over2} \hbar\left\{ \sum_{i=0}^{d}  k_i \; \Delta a_{i/2}  \; \Omega^{d+1-i} \right\}
+  {1\over2} \hbar  \; \Delta a_{(d+1)/2} \; \ln(\Omega^2/\mu^2) 
\nonumber\\&&\vphantom{\Bigg|}
+(\hbox{finite as } \Omega\to\infty).
\end{eqnarray}
Therefore, if the first $d+2$ Seeley--DeWitt coefficients, [from 0 to $(d+1)/2$], are equal, so that the differences are zero, 
$\Delta a_0 = \Delta a_{1/2} = \dots = \Delta a_{(d+1)/2} = 0$, then we have:
\begin{equation}
\Delta (\hbox{Casimir energy})_\mathrm{erfc}  =  (\hbox{finite as } \Omega\to\infty).
\end{equation}
\emph{We can now safely take the limit as  the cutoff is removed }($\Omega\to\infty$).
We have:
\begin{theorem}[Casimir energy differences]{\null \ \  \\}
If in $d$ space dimensions we compare two systems where the first $d+2$ Seeley--DeWitt coefficients are equal,
\begin{equation}
\Delta a_0 = \Delta a_{1/2} = \dots = \Delta a_{(d+1)/2} = 0, 
\end{equation}
then the erfc-regulated Casimir energy is finite as the regulator is removed:
\begin{equation}
\Delta (\hbox{Casimir energy}) =  (\hbox{finite}).
\end{equation}
(No renormalization is required.)
\end{theorem}
\noindent
This is a very nice mathematical theorem, but how relevant is it to real world physics?
Just how general is this phenomenon?

%=================================================================
\section{Equality between leading Seeley--DeWitt coefficients}
%=================================================================
Perhaps unexpectedly, there are \emph{very many} physically interesting situations where the (first few) Seeley--DeWitt coefficients are equal.
The pre-eminent cases are these (see for instance reference~\cite{Blau:1988} for a closely related discussion in terms of zeta-function regularization, more on this topic below): 
\begin{itemize}
\itemsep-1pt
\item Parallel plates.
\item Thin spherical shells.
\end{itemize}
In both of these cases an infra-red  (IR) regulator is needed at large distances, (effectively, put the entire system in some fixed large but finite size box or conducting sphere), and some subtle thought is still required.
But much more radically, take any collection of perfect conductors, and move them around relative to each other, (without distorting their shapes and/or volumes). 
\begin{itemize}
\itemsep-1pt
\item Then the change in Casimir energy is finite (no renormalization required).
\item Then the Casimir forces are finite (no renormalization required). 
\end{itemize}
(Subsequently, we shall show that similar comments can also be made for both imperfect conductors and dielectrics.)
To establish these results we note that for a region $\V$ with boundary $\partial \V$ we have the quite standard results that
\begin{eqnarray}
a_0 &\propto& \int_\V 1 \; \sqrt{g_d} \; \d^d x =(\hbox{volume});
\\
a_{1/2} &\propto&  \int_{\partial \V} 1 \;  \sqrt{g_{d-1}} \;\d^{d-1} x = (\hbox{surface area});
\\
a_1 &\propto&  \int_\V  \{R,V \}\;  \sqrt{g_d} \;\d^d x  +\int_{\partial \V} \{K\} \;  \sqrt{g_{d-1}}\;\d^{d-1} x;
\\
a_{3/2} &\propto& \int_{\partial \V} \{R,V, K^2, K_{ij}K^{ij}\} \;  \sqrt{g_{d-1}}\;\d^{d-1} x;
\end{eqnarray}
and
\begin{eqnarray}
a_2 &\propto &  
\int_\V  \{R^2,V^2,RV,\nabla^2 R,\nabla^2 V, R_{ab}R^{ab}, R_{abcd}R^{abcd} \}\;  \sqrt{g_d}\;\d^d x  
\nonumber\\
&&+\int_{\partial \V} \{R_{;n},V_{;n}, K_{ii:jj}, K_{ij:ij}, V K, K^3, \tr(K^2) K, \tr(K^3) \} \; \sqrt{g_{d-1}}\; \d^{d-1} x
\nonumber\\
&& +\int_{\partial \V} \{ RK, g^{ij} R_{ninj} K, R_{ninj} K^{ij}, g^{ik} R_{ijkl} K^{jl}\} \;  \sqrt{g_{d-1}}\;\d^{d-1} x.
\end{eqnarray}
See for instance references~\cite{Plunien:1986,Bordag:2001,Blau:1988,Blau:1988b,Visser:2002}, and the more extensive results in references~\cite{Gilkey:1990, Gilkey:2000, User-manual}. 
Here the $\{\_,\_,\_\}$ notation denotes various species-dependent linear combinations of the relevant terms. 

For current purposes we only need qualitative information --- we do not need to know the specific values of any of the dimensionless coefficients.
(In principle there are also contributions to the $a_i$ from kinks and corners; see for instance reference~\cite{Abalo:2012}; but  let's stay with smooth boundaries for now.)
Above we have retained terms due to both intrinsic and extrinsic curvature, plus a scalar potential $V(x)$. One could in principle obtain even more terms from non-zero background electromagnetic or gauge fields, but the terms retained above are entirely sufficient for current purposes. 

%=================================================================
\subsection{Parallel plate geometries}
%=================================================================
Working with gauge-fixed QED in flat spacetime with flat boundaries we note: (1) the scalar potential is zero, $V=0$, (2) the bulk Riemann tensor is zero, and (3) the extrinsic curvature of the flat plates is zero. Therefore the Seeley-DeWitt coefficients simplify to:
\begin{eqnarray}
a_0 &\propto&  (\hbox{volume});
\\
a_{1/2} &\propto&  (\hbox{surface area});
\\
a_1 &=&  0;
\\
a_{3/2} &=& 0;
\\
a_2 &=&  0.
\end{eqnarray}
So for a finite Casimir energy difference one just needs to keep volume and surface area fixed.
For example: 
Apply periodic boundary conditions in the $d-1$ spatial directions parallel to the plates, and
apply conducting box boundary conditions in the remaining spatial direction perpendicular to the plates. 

Physically this means you put the Casimir plates inside a big box, of fixed size, with two faces parallel to the plates. Then consider the situation where one varies the distance between the Casimir plates while keeping the size of the big box (the infra-red [IR] regulator) fixed.  From the above argument, and with no further calculation being required, we can at least deduce that the Casimir energy difference (and so the Casimir force between the plates) is finite.

%=================================================================
\subsection{Hollow spheres}
%=================================================================
Let us now consider gauge-fixed QED in flat spacetime with thin spherical boundaries. 
The idea is to understand as much as we can regarding Boyer's calculation~\cite{Boyer:1968}, but without explicit computation.
(We shall of course work in 3+1 dimensions.)

%=================================================================
\subsubsection{Step I (QED in flat spacetime)}
%=================================================================
\noindent
Using only the fact that we are working with QED ($V=0$) in flat spacetime (Riemann tensor zero):
\begin{eqnarray}
a_0 &\propto& (\hbox{volume});
\\
a_{1/2} &\propto&  (\hbox{surface area});
\\
a_1 &\propto&  \int_{\partial \V} \{K\} \; \sqrt{g_2}\; \d^{2} x;
\\
a_{3/2} &\propto& \int_{\partial \V} \{K^2, K_{ij}K^{ij}\} \;  \sqrt{g_2}\;\d^{2} x;
\\
a_2 &\propto&  \int_{\partial \V} \{ g^{ij} g^{kl} K_{ij:kl}, K^{ij}{}_{:ij}, K^3, \tr(K^2) K, \tr(K^3)\} \;  \sqrt{g_2}\; \d^{2} x.
\end{eqnarray}
Since the extrinsic curvature of the spherical shells is now non-zero, $K\neq0$, keeping control of the higher-order Seeley--DeWitt coefficients $a_i$ is now a little trickier.

%=================================================================
\subsubsection{Step II (thin boundaries)}
%=================================================================
\noindent
As long as the boundaries are thin, then for the extrinsic curvatures $K_\mathrm{inside} = - K_\mathrm{outside}$, 
leading to cancellations in both $a_1$ and $a_2$, which depend only on odd powers of extrinsic curvature. 
Similarly the thin boundaries take up zero volume, so the total volume is held fixed.
(The outermost boundary, the IR regulator,  is always held fixed.)

So the Seeley--DeWitt  coefficients simplify to:
\begin{eqnarray}
\Delta a_0 &\to&0;
\\
\Delta a_{1/2} &\propto&  \Delta(\hbox{surface area});
\\
\Delta a_1 &\to&0;
\\
\Delta a_{3/2} &\propto& \Delta \int_{\partial \V} \{K^2, K_{ij}K^{ij}\} \;  \sqrt{g_2}\;\d^{2} x;
\\
\Delta a_2 &\to& 0.
\end{eqnarray}
Note that we still need to worry about $a_{1/2}$ and $a_{3/2}$.

%=================================================================
\subsubsection{Step III (rescaling --- conformal invariance)}
%=================================================================
\noindent
As long as the inner boundaries for the two situations we are considering are simply rescaled versions of each other,
then the quantity $\int_{\partial\V} K K \;\sqrt{g_2} \;\d^2 x$ is scale invariant,
thus leading to a cancellation in $a_{3/2}$.
(The outermost boundary, the IR regulator, is always held fixed.)
Then:
\begin{eqnarray}
\Delta a_0 &\to&0;
\\
\Delta a_{1/2} &\propto&  \Delta(\hbox{surface area});
\\
\Delta a_1 &\to&0;
\\
\Delta a_{3/2} &\to& 0;
\\
\Delta a_2 &\to& 0.
\end{eqnarray}
Note we still have to deal with $\Delta a_{1/2}$. 

%=================================================================
\subsubsection{Step IV (TE and TM modes)}
%=================================================================
\noindent
In spherical symmetry, one can easily define TE and TM modes.
Note that they have equal and opposite contributions to $a_{1/2}$,
again leading to a cancellation in $a_{1/2}$.
(The outermost boundary is always held fixed.)
Then:
\begin{eqnarray}
\Delta a_0 &\to&0;
\\
\Delta a_{1/2} &\to&0;
\\
\Delta a_1 &\to&0;
\\
\Delta a_{3/2} &\to& 0;
\\
\Delta a_2 &\to& 0.
\end{eqnarray}
This finally is enough to guarantee finiteness of the Casimir energy difference. 

%=================================================================
\subsubsection{Step V (finiteness)}
%=================================================================
\noindent
From the above we have
\begin{equation}
\Delta (\hbox{Casimir energy}) = (\hbox{finite without renormalization}).
\end{equation}
This observation underlies the otherwise quite ``miraculous cancellations'' in Boyer's calculation
 of the Casimir energy of a hollow sphere~\cite{Boyer:1968}.
Comparing two hollow spheres of radius $a$ and $b$; 
and letting the IR regulator (which is the same for each sphere) move out to infinity:
\begin{equation}
\Delta (\hbox{Casimir energy}) =  \hbar \, c  \; B \left({1\over a} - {1\over b}\right).
\end{equation}
All one needs to do is ``merely'' to calculate the numerical coefficient $B$, 
which is now (thanks to our argument above) guaranteed to be finite.
(Boyer finds $B\approx  +0.04616...$, see reference~\cite{Boyer:1968} and follow-up discussion in references~\cite{Balian:1977} and~\cite{Milton:1978}.)
It is particularly important to realize that if one has somehow determined that
$\Delta (\hbox{Casimir energy}) = (\hbox{finite without renormalization})$,
then
\begin{equation}
\Delta (\hbox{Casimir energy}) = {1\over2}\, \hbar \;\sum\nolimits_{\{\emph{\hbox{any regular resummation technique}}\}} \; (\omega_n - (\omega_*)_n).
\end{equation}
Boyer uses Riesz resummation, (the so-called ``Riesz means''), the use of which is justified only in hindsight. If~you know the answer you want is finite, then any of the standard ``regular'' resummation techniques will do~\cite{Divergent}.
In contrast, if you don't know beforehand that the answer you want is finite, then blindly calculating
\begin{equation}
\sum_n (\omega_n - (\omega_*)_n)
\end{equation}
is simply asking for trouble.

%=================================================================
\subsection{Arbitrary arrangement of fixed-shape fixed-volume perfect conductors}
%=================================================================
Consider now any collection of fixed-shape fixed-volume perfect conductors in 3+1 dimensions. We are working with gauge-fixed QED ($V=0$) in flat spacetime (Riemann tensor zero). Then:
\begin{eqnarray}
a_0 &\propto& (\hbox{volume});
\\
a_{1/2} &\propto&  (\hbox{surface area});
\\
a_1 &\propto&  \int_{\partial \V} \{K\} \;  \sqrt{g_2}\; \d^{2} x;
\\
a_{3/2} &\propto& \int_{\partial \V} \{K^2, K_{ij}K^{ij}\} \;  \sqrt{g_2}\;\d^{2} x;
\\
a_2 &\propto&  \int_{\partial \V} \{ g^{ij} g^{kl} K_{ij:kl}, K^{ij}{}_{:ij}, K^3, \tr(K^2) K, \tr(K^3)\} \;  \sqrt{g_2}\;\d^{2} x.
\end{eqnarray}
But fixed-shape and fixed-volume implies fixed extrinsic curvature, so all the $\Delta a_i \equiv 0$.
That is:
\begin{itemize}
\itemsep-3pt
\item Take any collection of perfect conductors. \\
Move them around relative to each other.\\
(Without distorting their shapes and/or volumes.)
\item Then the change in Casimir energy, and so the Casimir forces, \\
are finite without renormalization.
\end{itemize}
We shall subsequently see how to generalize this result to imperfect conductors and/or dielectrics.

%=================================================================
\section{Zeta function techniques: Regularization}
%=================================================================
Let us now relate the discussion above to zeta function techniques, see for instance~\cite{Blau:1988}. 
The key observation is to write
\begin{eqnarray}
\zeta(s) &=& \sum_n (\omega_n/\mu)^{-2s} =  {1\over\Gamma(s)} \int_0^\infty \d u \; u^{-1+s} \; \sum_n  \exp(-u \omega^2/\mu^2 ) 
\nonumber\\&&
= {1\over\Gamma(s)} \int_0^\infty \d u \; u^{-1+s} \;K(u/\mu^2 ).
\end{eqnarray}
Here $\mu$ is some convenient scale introduced to keep the zeta function dimensionless.
Then, when comparing two situations for which the first $d+2$ Seeley-DeWitt coefficients are equal we have 
\begin{equation}
\zeta(s) - \zeta_*(s) = \hbox{(analytic function of $s$ without poles for $s\geq 1/2$).}
\end{equation}
In this situation we can unambiguously write
\begin{equation}
\Delta (\hbox{Casimir energy}) = {1\over2}\, \hbar\,\mu \left\{ \zeta(-1/2) - \zeta_*(-1/2) \right\},
\end{equation}
which is unambiguously finite (and in fact independent of $\mu$) without any need for renormalization.
However, if any of the first $d+2$ Seeley-DeWitt coefficients differ the situation is more complicated; one has to renormalize not just regularize. More on this point below.

%=================================================================
\section{The use of reference models}
%=================================================================
Consider now a non-zero potential ($V\neq 0$), in flat spacetime (Riemann tensor zero), with periodic boundary conditions in space (so that there is no boundary).
We have:
\begin{eqnarray}
a_0 &\propto& (\hbox{volume});
\\
a_{1/2} &=&  0;
\\
a_1 &\propto&  \int_\V  \{V \}\; \sqrt{g_d}\;\d^d x;
\\
a_{3/2} &=& 0;
\\
a_2 &\propto&  \int_\V  \{V^2 \}\; \sqrt{g_d}\;\d^d x.
\end{eqnarray}
So for finiteness we ``\emph{just}'' need to keep $a_0$, $a_1$, and $a_2$ fixed.

%=================================================================
\subsection{1+1 dimensions}
%=================================================================
In (1+1) dimensions let us define the spatial average (assumed non-negative for simplicity)
\begin{equation}
\overline V = {\int_0^L V(x) \; \d x\over L}.
\end{equation}
Let us now compare the two situations:
\begin{itemize}
\itemsep-3pt
\item $D =- \nabla^2 + V(x)$; \qquad eigenvalues $\omega_n^2$.
\item $\overline D = -\nabla^2 + \overline V$; \qquad \quad \, eigenvalues $\overline\omega_n^2$.
\end{itemize}
Then:
\begin{equation}
\sum_n \left\{ \omega_n \; \erfc(\omega_n/\Omega) - \overline\omega_n \; \erfc(\overline\omega_n/\Omega) \right\}
 =(\hbox{finite as } \Omega\to\infty).
\end{equation}
That is:
\begin{equation}
\hbox{(Casimir energy of }D) - \hbox{(Casimir energy of }\overline D)  =(\hbox{finite}).
\end{equation}
In fact in this situation the reference eigenvalues $\bar\omega_n$ can be written down explicitly as
\begin{equation}
\overline\omega_n =  \sqrt{ {(2\pi n)^2\over L^2} + \overline V}.
\end{equation} 
Thence
\begin{equation}
\sum_n \left( \omega_n -  \sqrt{ {(2\pi n)^2\over L^2} + \overline V}\,\right)  =(\hbox{finite})  .
\end{equation} 
The $\omega_n$ depend on $V(x)$ and can be quite messy; but the difference between the $\omega_n$ and the reference problem $\overline\omega_n$ is however guaranteed to be well behaved.

%=================================================================
\subsection{3+1 dimensions}
%=================================================================
In (3+1) dimensions define  the two spatially averaged quantities: 
\begin{equation}
\overline V = {\int_\V V(x) \;\d^3 x\over \hbox{volume($\V$)}}; 
\qquad
\overline{V^2} = {\int_\V V(x)^2 \;\d^3 x\over \hbox{volume($\V$)}}.
\end{equation}
Now solve
\begin{equation}
m_1^2 + m_2^2 = 2 \overline V; \qquad m_1^4+m_2^4 = 2 \overline{V^2},
\end{equation}
to determine two parameters $m_1$ and $m_2$.
Then compare these three situations:
\begin{itemize}
\itemsep-3pt
\item $D = -\nabla^2 + V(x)$; \qquad eigenvalues $\omega_n^2$.
\item $\overline{D_1} = -\nabla^2 + m_1^2$; \qquad \,eigenvalues $\overline{(\omega_1)}_n^2$.
\item $\overline{D_2} = -\nabla^2 + m_2^2$; \qquad \,eigenvalues $\overline{(\omega_2)}_n^2$.
\end{itemize}
\noindent
(We assume $m_i^2>0$ for simplicity.)
We see that:
\begin{eqnarray}
&&\sum_n \left\{ \omega_n \; \erfc(\omega_n/\Omega) 
- {1\over2}\;\overline{(\omega_1)}_n \; \erfc\left(\,\overline{(\omega_1)}_n/\Omega\right)
- {1\over2}\;\overline{(\omega_2)}_n \; \erfc\left(\,\overline{(\omega_2)}_n/\Omega\right) \right\}
\nonumber\\&& \qquad\qquad\qquad
 =(\hbox{finite as } \Omega\to\infty). \qquad
\end{eqnarray}
This implies that physically:
\begin{eqnarray}
&&\hbox{(Casimir energy of }D) - {1\over2}\; \left(\hbox{Casimir energy of }\overline{D_1}\right)  - {1\over2}\; \left(\hbox{Casimir energy of }\overline{D_2}\right)  
\nonumber\\&& \qquad\qquad\qquad\qquad
=(\hbox{finite}).
\end{eqnarray}
While this argument does not calculate the finite piece for you, it at least guarantees that you are looking for a finite answer. 
(And since the answer you are looking for us guaranteed finite, almost any way of manipulating the series and thereby getting to that finite answer will give the correct answer.)

But the two comparison models $\overline{D_1}$ and $\overline{D_2}$ are now sufficiently simple that one can invoke analytic techniques, (zeta functions and the like, see for instance references~\cite{Blau:1988, Blau:1988b}),
and simply define
\begin{eqnarray}
\hbox{(Casimir energy of }D) &=& {1\over2} \left(\hbox{Casimir energy of }\overline{D_1}\right)  + {1\over2} \left(\hbox{Casimir energy of }\overline{D_2}\right)  
\nonumber\\&& \qquad
+(\hbox{finite}).
\end{eqnarray}
Of course this does not calculate the ``finite piece'' for you, but it gives you some confidence regarding what to aim for before you start calculating.

%=================================================================
\section{What if Casimir energy differences are not naively finite?}
%=================================================================

Now there are certainly (mathematical) situations where the $\Delta a_i\neq 0$ and the Casimir energy difference is not naively finite.  
This merely means one has to be more careful thinking about the physics. (One might need to \emph{renormalize}, not just \emph{regularize}, or one might have good physics reasons to keep the regularization parameter finite, and not drive it to infinity.) For instance: 
\begin{itemize}
\itemsep-3pt
\item Real metals and real dielectrics are transparent in the UV.
\item The UV cutoff $\Omega$ is then merely a stand-in for all the complicated physics. 
\end{itemize}
It is important to emphasize that for real metals and real dielectrics the cutoff represents real physics. See for instance the discussion in references~\cite{Schwinger:1977, Milton:1978, Milton:2004}, and compare with the discussion in \cite{Carlson:1996, Carlson:1997, MolinaParis:1997, Liberati:2000}. Note that the discussion regarding real metals and real dielectrics has often lead to some considerable disagreement regarding interpretation~\cite{Jaffe:2005, Graham:2002, Graham:2003}. 
(My own view, as should be clear from the current article, is that physical Casimir energies are ultimately determined by looking at physical differences in zero-point energies, summed over all relevant modes.)

%=================================================================
\subsection{A very general class of cutoff functions}
%=================================================================

Let us write a very general class of cutoff functions as follows:
\begin{equation}
f\left(\omega\over\Omega\right) = \int_0^\infty g(\xi) \;\erfc\left(\omega\over\xi\Omega\right)  \; \d \xi.
\end{equation}
Here we demand
\begin{equation}
g(\xi)\geq 0; \qquad
\int_0^\infty g(\xi) \;\d \xi = 1;   \qquad 
\langle \xi^i\rangle = \int_0^\infty g(\xi) \; \xi^i \;\d \xi <\infty;  \qquad i\in\{1,2,\dots,d\}.
\end{equation}
So $g(\xi)$ is non-negative, normalized, and its first few moments are finite.
Note $f(0)=1$, while $f(\infty)=0$, and $f(\omega/\Omega)$ is monotone decreasing.
To see just how general this class of cutoff functions is, we proceed by noting that
\begin{equation}
\erfc\left(\omega\over\xi\Omega\right) =  
{2\over\Omega\xi\sqrt{\pi}} 
\int_\omega^\infty \exp\left(-{x^2\over\Omega^2\xi^2}\right) \d x.
\end{equation}
So differentiating $f(\_)$ we see
\begin{equation}
f'\left(\omega\over\Omega\right) = -{2\over\sqrt{\pi}}\int_0^\infty {g(\xi)\over\xi} \; \; \exp\left(-{\omega^2\over\Omega^2\xi^2}\right)    \; \d \xi.
\end{equation}
Substituting $\chi=1/\xi^2$ we obtain
\begin{equation}
f'\left(\omega\over\Omega\right) = {1\over\sqrt{\pi}}\int_0^\infty {g(\chi^{-1/2})\over\chi} \; \; \exp\left(-{\omega^2\over\Omega^2}\; \chi\right)    \; \d \chi.
\end{equation}
But this is just the Laplace transform of ${g(\chi^{-1/2})/\chi}$,  evaluated at the point $s=\omega^2/\Omega^2$. Consequently, as long as the inverse Laplace transform of $f'(s^{1/2} )$ exists, which is a relatively mild condition on the cutoff function $f(s^{1/2})$,  then we can determine $g(\xi)$ in terms of $f(\omega/\Omega)$. 

Indeed, there is a little-known formal algorithm due to Post~\cite{Post}, see also the discussion by Bryan~\cite{Bryan}, and which is  further discussed in reference~\cite{Sotiriou:2011}, that allows for formal inversion of Laplace transforms by taking arbitrarily high derivatives.
Specifically, if $G(s)$ is the Laplace transform of $g(z)$ then

\begin{equation}
g(z) = \lim_{n\to\infty}  {(-1)^n\over n!} \; \left({n\over z}\right)^{n+1}  \; G^{(n)} \left({n\over z}\right).
\end{equation}
This algorithm may not always be particularly practical, since one needs arbitrarily high derivatives. But even if not always practical, the mere existence of this algorithm settles an important issue of principle --- knowledge of the cutoff $f(\omega/\Omega)$ in principle allows one to reconstruct an equivalent weighting function $g(\xi)$.
The point is that almost any physically reasonable cutoff function $f(\omega/\Omega)$ can be cast in this ``weighted integral over erf-functions'' form. (In particular we could rephrase all of the preceding discussion concerning erf-regularization in terms of this more general $f$-regularization, but when $\Delta a_i=0$ nothing new is obtained. It is only when  the $\Delta a_i\neq0$ that general $f$-regularization becomes at all interesting.)

%=================================================================
\subsection{$f$-regularized Casimir energy}
%=================================================================
Let us now consider a generic regularized sum of eigen-frequencies:
\begin{equation}
\sum_n \omega_n \; f\left(\omega_n\over\Omega\right).
\end{equation}
Then our previous result
\begin{equation}
\sum_n \omega_n \; \erfc(\omega_n/\Omega) =   \left\{ \sum_{i=0}^{d}  k_i \; a_{i/2}  \; \Omega^{d+1-i} \right\}
+ a_{(d+1)/2} \; \ln(\Omega^2/\mu^2) 
+(\hbox{finite as } \Omega\to\infty),
\end{equation}
becomes
\begin{eqnarray}
\sum_n \omega_n \;  f\left(\omega_n\over\Omega\right) &=&   \left\{ \sum_{i=0}^{d}  k_i \left(\int_0^\infty g(\xi) \;\xi^{d+1-i} \;\d \xi\right) \; a_{i/2}  \; \Omega^{d+1-i} \right\}
\nonumber\\
&&
+ a_{(d+1)/2} \; \left\{ \ln(\Omega^2/\mu^2) + 2 \int_0^\infty g(\xi) \;\ln\xi \;\d \xi\right\} 
\nonumber\\&&\vphantom{\Bigg|}
+(\hbox{finite as } \Omega\to\infty).\qquad
\end{eqnarray}
That is:
\begin{eqnarray}
\sum_n \omega_n \;  f\left(\omega_n\over\Omega\right) &=&   \left\{ \sum_{i=0}^{d}  k_i \;\langle\xi^{d+1-i}\rangle \; a_{i/2}  \; \Omega^{d+1-i} \right\}
\nonumber\\
&&
+ a_{(d+1)/2} \; \left\{ \ln(\Omega^2/\mu^2) + 2 \,\langle\ln\xi\rangle\right\} 
+(\hbox{finite as } \Omega\to\infty).\qquad
\end{eqnarray}
The integrals over $g(\xi)$ can now be absorbed into redefining the dimensionless constants $k_i$ in a $f$-dependent (and hence $g$-dependent) manner, and absorbing the $\langle\ln\xi\rangle$ term into the finite piece. 
That is, we have:
\begin{theorem}[Physical cutoff]\ \\
For a general physical cutoff $f(\omega/\Omega)$ one has
\begin{equation}
\sum_n \omega_n \;  f\left(\omega_n\over\Omega\right)
 =   \left\{ \sum_{i=0}^{d}  [k(f)]_i \; \;a_{i/2}  \;\; \Omega^{d+1-i} \right\}
+ a_{(d+1)/2} \;\; \ln(\Omega^2/\mu^2) 
+(\hbox{\emph{finite as }} \Omega\to\infty).
\end{equation}
The $[k(f)]_i = k_i \, \langle\xi^{d+1-i}\rangle$ are dimensionless phenomenological parameters that depend on the detailed physics of the specific cutoff function $f(\omega/\Omega)$. 
\emph{The $\Omega$ dependence represents real physics. Live with it!} 
\hfill{$\Box$}
\end{theorem}
\noindent
Now define the physically regularized Casimir energy. 
For a general cutoff $f(\omega/\Omega)$ we set
\begin{equation}
(\hbox{Casimir energy)}_{f} = {1\over2} \hbar\; \left(\sum_n \omega_n \;  f\left(\omega_n\over\Omega\right) \right).
\end{equation}
Then we deduce:
\begin{theorem}[Physically regularized Casimir energy]\ \\
For a general physical cutoff $f(\omega/\Omega)$ one has
\begin{eqnarray}
(\hbox{\emph{Casimir energy)}}_{f}
&=&  {1\over2} \hbar \left\{ \sum_{i=0}^{d}  [k(f)]_i \; a_{i/2}  \; \Omega^{d+1-i} \right\}
+ {1\over2} \hbar \; a_{(d+1)/2} \; \ln(\Omega^2/\mu^2) 
\nonumber\\&&\qquad\vphantom{\Bigg|}
+(\hbox{\emph{finite as }} \Omega\to\infty).
\end{eqnarray}
The $[k(f)]_i = k_i \, \langle\xi^{d+1-i}\rangle$ are dimensionless phenomenological parameters that depend on the 
detailed physics of the specific cutoff function $f(\omega/\Omega)$. 
\emph{The $\Omega$ dependence represents real physics.  Live with it!}
\hfill{$\Box$}
\end{theorem}
\noindent
Part of the reason it was never worthwhile to keep explicit track of the $k_i$ is that, once the $f$-cutoff is introduced, the $k_i$ would in any case be replaced by the purely phenomenological and cutoff dependent coefficients $[k(f)]_i$.

Furthermore, if we compare two systems, and if the first $d+2$ of the $\Delta a_i$ are zero, then the cutoff dependence drops out of the calculation. That is, even for imperfect conductors and dielectrics, if one is comparing two situations where the conductors/dielectrics have merely been moved around, (without changing shape and/or volume), then the difference in Casimir energies (and so the Casimir forces) are guaranteed finite  (and cutoff independent).

%=================================================================
\section{Enforcing finiteness?}
%=================================================================
Can one \emph{force} the Casimir energy difference to be finite?
Suppose that by hook or by crook one can find a number $m$ of ``simple'' tractable problems $\overline{D_i}$ such that
\begin{equation}
a_{i/2}(D) = \sum_{i=1}^m p_i \; a_{i/2}(\overline{D_i});   \qquad \sum_{i=1}^m p_i = 1;  \qquad i\in \{0,1,2, \dots, d+1\}.
\end{equation}
Then it is certainly safe to say
\begin{equation}
\hbox{(Casimir energy of }D) - \sum_{i=1}^m p_i \, \left(\hbox{Casimir energy of }\overline{D_i}\right)
=(\hbox{finite}).
\end{equation}
Of course this does not calculate the ``finite piece'' for you, but it gives you some confidence regarding what to aim for before you start calculating. 
More formally, if the $\overline{D_i}$ are sufficiently simple one might apply analytic techniques (such as zeta functions~\cite{Blau:1988, Blau:1988b} or the like) to argue that it might make sense to define:
\begin{equation}
\hbox{(Casimir energy of }D) = \sum_{i=1}^m p_i \, \left(\hbox{Casimir energy of }\overline{D_i}\right)
+(\hbox{finite}).
\end{equation}
A somewhat safer statement is to compare two systems and assert
\begin{equation}
\Delta \hbox{(Casimir energy of }D) =\sum_{i=1}^m \, p_i \, \Delta(\hbox{Casimir energy of }\overline{D_i})  
+(\hbox{finite}).
\end{equation}
Only if the two sets of ``reference problems'' $\overline{D_i}$ are the same, (or at the very least have the same weighted sum of Seeley-DeWitt coefficients  $\sum_{i=1}^m p_i\,a_j\left[\,\overline{D_i}\,\right]$), does this process make any real physical sense, in which case it reduces to our previous result
\begin{equation}
\Delta \hbox{(Casimir energy of }D) =  (\hbox{finite}).
\end{equation}
Otherwise the sum
\begin{equation}
\sum_{i=1}^m p_i\;\Delta(\hbox{Casimir energy of }\overline{D_i}),
\end{equation}
while analytically continued to be finite, is purely formal. It need not be a physical energy difference.
In short, one should seek at all times to calculate Casimir energy differences between clearly defined and specified physical systems. 
This might, at a pinch, involve differences between linear combinations of physical systems, but to get a physically meaningful Casimir energy one must either enforce $\Delta\left(\sum_{i=1}^m p_i\, a_j\left[\,\overline{D_i}\,\right]\right) =0$, or develop an explicit physical model for the cutoff function $f(\omega/\Omega)$.

%=================================================================
\section{Zeta function techniques: Renormalization}
%=================================================================
With all the discussion above now under control, let us now consider zeta function techniques in situations where \emph{renormalization} rather than \emph{regularization} is needed. (See for instance~\cite{Blau:1988}.)
Again the key observation is to write
\begin{eqnarray}
\zeta(s) &=& \sum_n (\omega_n/\mu)^{-2s} =  {1\over\Gamma(s)} \int_0^\infty \d u \; u^{-1+s} \; \sum_n  \exp(-u \omega^2/\mu^2 ) 
\nonumber\\
&=& {1\over\Gamma(s)} \int_0^\infty \d u \; u^{-1+s} \;K(u/\mu^2 ).
\end{eqnarray}
Here $\mu$ is again some convenient scale introduced to keep the zeta function dimensionless.
Then, in terms of the first $d+2$ Seeley-DeWitt coefficients  we have~\cite{Blau:1988}
\begin{equation}
\zeta(s) = {1\over\Gamma(s) \; (4\pi)^{d/2}} \; \sum_{n=0}^{\infty}   {a_{n/2} \; \mu^{d-2n} \over s-[d/2-n]}   + \hbox{(entire analytic function of $s$).}
\end{equation}
Note the presence of poles at $d/2-n$, for $n\in N$.  To develop a zeta-function definition of the Casimir energy we need to analytically continue from large positive $s$, specifically $s>d/2$, where the zeta function is guaranteed to converge, to $s\to -1/2$.
So only some of the poles are relevant for our purposes, namely the poles in the range $[-1/2, d/2]$. That is, we might as well write
\begin{equation}
\zeta(s) = {1\over\Gamma(s) \; (4\pi)^{d/2}} \; \sum_{n=0}^{(d+1)/2}   {a_{n/2} \; \mu^{d-2n} \over s-[d/2-n]}   
+ \hbox{(analytic function of $s$ for $s>-1$).}
\end{equation}
To \emph{define} the Casimir energy we now have two choices:
\begin{itemize}
\item The mathematically slick (but physically maybe not entirely reliable) trick of simply discarding the poles and using the principal part of the zeta function:
\begin{equation}
E_{Casimir} = {1\over2} \, \hbar \mu \; PP\{ \zeta(-1/2) \} =  
{1\over4} \, \hbar \mu \lim_{\epsilon\to 0}\left\{ \zeta(-1/2+\epsilon) + \zeta(-1/2-\epsilon) \right\}.
\end{equation}
This procedure (advocated by myself and co-authors in reference~\cite{Blau:1988}) is certainly mathematically efficient, but maybe a bit too slick in terms of hiding key parts of the physics. (Basically one is dodging into the complex $s$-plane to avoid looking at the poles in $s\in[0,d/2]$, dealing with the pole at $s=-1/2$ via the principal part prescription.)

\item The physically more reasonable alternative is to remain on the real $s$-line, not discard the poles one is passing over, and introduce a number of renormalization-induced phenomenological parameters to write:
\begin{eqnarray}
E_{Casimir} 
&=& 
\sum_{n=0}^{(d+1)/2}   K_n \; a_{n/2} + {1\over2} \, \hbar \mu \; PP\{ \zeta(-1/2) \} \\
&=&   
\sum_{n=0}^{(d+1)/2}   K_n \; a_{n/2} +
{1\over4} \, \hbar \mu \lim_{\epsilon\to 0}\left\{ \zeta(-1/2+\epsilon) + \zeta(-1/2-\epsilon) \right\}.
\end{eqnarray}
Setting up the zeta-function calculation in this way has the advantage of both matching the results obtained by other regularization and renormalization techniques, especially the physically based erfc and f-erfc regularization and renormalization techniques discussed above, and making it utterly clear exactly where the $\sum_{n=0}^{(d+1)/2}   K_n \; a_{n/2}$ terms are coming from and why they are physically necessary. 
\end{itemize}
As two examples of this construction consider the following:
\begin{itemize}
\item When comparing two situations for which the first $d+2$ Seeley-DeWitt coefficients are equal we have 
\begin{equation}
\Delta(E_{Casimir}) = {1\over2} \, \hbar \mu \; \Delta (\zeta(-1/2)).
\end{equation}
Because by assumption $\Delta a_{(d+1)/2}=0$ we can dispense with the $PP$ principal part prescription.
We thus recover the results of our earlier discussion --- in this situation regularization (not renormalization) is all that is needed. 

\item
Now consider a \emph{single} system for which $ a_{(d+1)/2}=0$, but the $a_n\neq 0$ for $n\in[0,d/2]$. 
Such systems certainly exist, see for instance the polyhedral systems discussed in reference~\cite{Abalo:2012}.
Because by assumption $a_{(d+1)/2}=0$ we can dispense with the $PP$ principal part prescription.
Then we have two possible definitions for the zeta-function renormalized Casimir energy:
\begin{equation}
E_{Casimir} = {1\over2} \, \hbar \mu \;\zeta(-1/2),
\label{E:kim}
\end{equation}
\emph{versus}
\begin{equation}
E_{Casimir} = 
\sum_{n=0}^{d/2}   K_n \; a_{n/2} + {1\over2} \, \hbar \mu \zeta(-1/2).
\label{E:not-kim}
\end{equation}
Which of these two options is physically more reasonable? The authors of reference~\cite{Abalo:2012} choose equation (\ref{E:kim}) thereby discarding the known infinities that certainly do show up in the naive mode sum $\sum_n \omega_n$. 
In fact the authors of reference~\cite{Abalo:2012}  explicitly verify that the known infinities proportional to the $a_n$ for $n\in [0,d/2]$  show up in their calculation, but then simply discard these infinities. I would argue that discarding these infinities,
while mathematically slick is physically unmotivated and physically unnecessary --- one loses useful information. Accordingly proper renormalization along the lines of equation (\ref{E:not-kim}), with a number of phenomenological parameters $K_n$ for $n\in[0,d/2]$, keeps physically relevant information while being full compatible with other renormalization techniques. 
\end{itemize}
In short, zeta-function techniques are mathematically powerful and slick, but it is easy to unnecessarily discard useful information.
Zeta-function techniques have a tendency to renormalize certain phenomenological parameters to zero, losing key information in the process. Some care and delicacy in applying zeta-function techniques is required. (See reference~\cite{Liberati:2000} for other cautionary comments regarding over-enthusiastic renormalization to zero of physically interesting quantities.)

%=================================================================
\section{Conclusions}
%=================================================================
In $(d+1)$ dimensions, iff the first $d+2$ Seeley--DeWitt coefficients agree, 
\begin{equation}
\Delta a_0 = \Delta a_{1/2} = \dots \Delta a_{(d+1)/2} = 0, 
\end{equation}
then the difference in Casimir energies is guaranteed finite without any need for renormalization. 
This is an extremely useful thing to check before you start explicitly calculating. 
Furthermore we have seen that the $\erfc$ function, in the form $\erfc(\omega/\Omega)$, is a perhaps unexpectedly useful regulator
\begin{equation}
\erfc(0)=1; \qquad \erfc(\infty) = 0.
\end{equation}
More generally, any reasonably smooth monotone function  $f(\omega/\Omega)$ satisfying
\begin{equation}
f(0)=1; \qquad f(\infty) = 0;
\end{equation}
will do. (Roughly speaking, as long as $f(\omega/\Omega)$ has an inverse Laplace transform.) 
For real metals and real dielectrics, which become transparent in the UV, the cutoff is physical, and its influence on the Casimir energy is encoded in a small number of dimensionless parameters $[k(f)]_i$ and an overall cutoff scale $\Omega$. 
Various generalizations of this argument, (such as counting differences in the number of eigenstates, or calculating differences of sums of powers of eigenvalues), are also possible. Similar arguments, regarding differences in Seeley--DeWitt coefficients, can also be applied to the one-loop effective action~\cite{Visser:2002}. Finally, I should emphasise that I have not \emph{renormalized} anything anywhere in  this article, the worst I have done is to temporarily \emph{regularize} some infinite series, to allow some otherwise formal manipulations to be mathematically (and physically) well-defined. 

Regarding Casimir forces, the current discussion is enough to specify when Casimir forces (being based on adiabatic variation of position parameters) are finite under regularization without renormalization. The dynamical Casimir effect~\cite{Carlson:1996,Liberati:2000,Wilson:2011,Felicetti:2014}, involving a rapid change of positional and other parameters is considerably trickier. 
It would be difficult to draw broad conclusions regarding the dynamical Casimir effect based on the current article.

\vspace{10pt}
\hrule
\medskip
%------------------------------------------------
%\authorcontributions{single author}

\acknowledgments{
This research was supported by the Marsden Fund, and by a James Cook fellowship,
both administered by the Royal Society of New Zealand. 
The author would like to express a special thanks to the Mainz Institute for Theoretical Physics
 (MITP) for its hospitality and support.}

\noindent
{\sc Conflicts:} The author declares no conflict of interest.

\bigskip
\hrule

%------------------------------------------------------------------------------------------------------------------------------------------
%------------------------------------------------------------------------------------------------------------------------------------------
%------------------------------------------------------------------------------------------------------------------------------------------
%------------------------------------------------------------------------------------------------------------------------------------------

%------------------------------------------------------------------------------------------------------------------------------------------
%------------------------------------------------------------------------------------------------------------------------------------------

%%%%%%%%%%%%%%%%%%%%%%%%%%%%%%%%%%%%%%%%%%%%%%%%%
\end{document}